\documentstyle[art12]{article}
\begin{document}
\begin{titlepage}
\begin{flushright}
FERMILAB-CONF-95-287-T\\
\end{flushright}
\vspace{1cm}
\begin{center}
{\Large\bf CP Violation and Quantum Mechanics \\
in the $B$ System
\footnote{To appear in the Proceedings of the Moriond Workshop on
Electroweak Interactions and Unified Theories, Les Arcs, France March
1995}}\\
\vspace{1cm}

{\large
Boris Kayser}\\
\vspace{0.4cm}
{\it
Fermi National Accelerator Laboratory \\
Batavia, IL 60510 USA\\
and\\
National Science Fundation\\
4201 Wilson Blvd.,\\
Arlington, VA 22230, USA} \\

\end{center}

\begin{abstract}
  We discuss the testing of the Standard Model of CP violation, and
the search for CP-violating effects from beyond the Standard Model, in
$B$ decays.  We then focus on the quantum mechanics of the experiments
on CP violation to be performed at $B$ factories. These experiments
will
involve very pretty Einstein-Podolsky-Rosen correlations. We
show that the physics of these experiments can be understood without
invoking the ``collapse of the wave function," and without the
mysteries that sometimes accompany discussions of EPR effects.
\end{abstract}

\end{titlepage}
\section*{Introduction}

We are anticipating that future measurements of CP-violating  asymmetries
in $B$ decays will cleanly and incisively test the Standard Model (SM)
description of CP violation. Physics beyond  the SM could reveal itself
through failures of the SM predictions for these asymmetries. As we shall
see, the future experiments will provide a beautiful  example of the
workings of quantum mechanics.

We first discuss the $B$-system test of the SM of CP violation. What
quantities would one like to measure in order to carry out this test? How
can these quantities be measured cleanly? Which $B$ decay modes probe each
quantity? Finally, how can physics beyond the SM affect the CP-violating
effects to be studied?

We then turn to the quantum mechanics of the planned CP experiments at $B$
factories. These experiments will involve a very pretty modern example of
an Einstein--Podolsky--Rosen (EPR) correlation. We show that the
experiments can be understood through an approach based entirely on
amplitudes, rather than on wave functions. This approach is manifestly
covariant, and does not entail the somewhat mysterious  ``collapse of the
wave function" which is usually invoked to describe EPR effects.

\section*{CP Violation in the $B$ System}

According to the SM, CP violation is a consequence of the fact that in the
Cabibbo--Kobayashi--Maskawa (CKM) quark mixing matrix,

\begin{equation}
V=\left(\begin{array}{ccc}
V_{ud} & V_{us} & V_{ub} \\ V_{cd} & V_{cs} & V_{cb} \\
V_{td} & V_{ts} & V_{tb} \end{array} \right) \;,
\end{equation}
some of the elements are not real, but complex. CP-violating effects in $B$
decays can severely test this hypothesis by cleanly determining the phases
of various products of CKM elements. In principle, one would like to
determine all the independent phases of this kind. How may such independent
phases exist, and what are they?

To answer this question, we recall that the SM requires the CKM matrix $V$
to be unitary. This unitarity imposes, among other constraints, the
orthogonality conditions
\begin{equation}
\sum^3_{\alpha =1} V_{\alpha i}V^*_{\alpha j} = 0; \;\;ij=ds, sb, db \;.
\end{equation}
Here, $V_{\alpha i}$ is an element of $V$. For given $ij$, the
orthogonality condition (2) is conveniently pictured as the statement that
the individual terms in the condition form the sides of a {\it closed}
triangle (a ``unitarity triangle") in the complex plane. From the existing
information on the magnitudes of the various $V_{\alpha i}$, we expect that
the triangle for $ij=db$ (the ``$db$ unitarity triangle") will have sides
of comparable length, so that its interior angles $\alpha , \beta ,$ and
$\gamma$ may all be large. This triangle is shown schematically in Fig.~1.
By contrast, we expect that in the $sb$ unitarity triangle, the
$V_{us}V^*_{ub}$ side is only $\sim 0.02$ as long as the other two sides.
Thus, the angle opposite the $V_{us} V^*_{ub}$ side, which we shall call
$\epsilon,$ is \def\ltap{\;\raisebox{-.4ex}{\rlap{$\sim$}}
\raisebox{.4ex}{$<$}\;} $\!\!\ltap 0.02$ radians. Similarly, we expect that
in the $ds$ triangle, the $V_{td}V^*_{ts}$ side is only $\sim 0.002$ as
long as the other sides, so that the angle $\epsilon^\prime$ opposite this
short side is $\ltap 0.002$ radians. Suppose, now, that $\phi$ is the phase
of some convention-independent product of CKM elements. Then it can be
shown\cite{1} that
\begin{equation}
\phi =n_\alpha \alpha +\eta_\beta \beta +\eta_\epsilon
\epsilon+\eta_{\epsilon^\prime}
\epsilon^\prime \;,
\end{equation}
where $\eta_{\alpha , \beta ,\epsilon ,\epsilon^\prime}$ are integers. Thus,
the four angles $\alpha ,\beta ,\epsilon ,$ and $\epsilon^\prime$ in the
unitarity
triangles are the independent phases of all possible (convention-independent)
products of CKM elements. Since the phases of CKM products are
the quantities which will be determined by the experimental studies of CP
violation in the $B$ system, these studies may be thought of as probes of
these
four angles. Quite possibly, the angle $\epsilon^\prime$, which is at most a
few milliradians and leads to CP-violating effects which, correspondingly,
are at most a few parts per $10^3$, will prove to be beyond experimental
reach.
However, experiments which hopefully will determine the remaining
angles,
$\alpha ,\beta ,$ and $\epsilon$, are actively being developed.

Wolfenstein has introduced a very good $(\sim 3\%)$ approximation to the
CKM matrix\cite{2} in which $\epsilon =\epsilon^\prime =0$. In this
approximation, the only nontrivial independent phase angles are  $\alpha$
and $\beta$ in the $db$ triangle. For this reason, in the literature,
attention has been properly focussed on this triangle.

The goals of the experiments aimed at testing the SM of CP violation
through  studies of the $B$ system can be summarized in the following way:
First, to measure the four independent angles of the unitarity triangles,
or at  least three of them $(\alpha ,\beta ,\epsilon$). Attention will be
focussed first on the angles $\alpha$ and $\beta$, since these may both be
large. Secondly, to overconstrain the system as much as possible. To do so,
one can:  a) See if CP asymmetries in different decay modes, which all
yield the same angle (say, $\beta$) {\it if} the SM of CP violation is
correct, actually yield the same numerical result. (b) Measure
independently the angles $\alpha ,\beta ,$ and the dependent angle $\gamma$
in the $db$ triangle and see whether these three angles actually add up to
$\pi$. (c) Measure the lengths of the sides of the $db$ triangle (via
studies on non-CP-violating effects such as decay rates and neutral $B$
mixing), and then see whether the interior angles implied by these lengths
agree with those  inferred directly from CP-violating asymmetries. Needless
to say, overconstraining the system in these ways will enable one to test
whether the SM provides a consistent picture of CP-violating  phenomena, or
leads to inconsistencies which point to physics beyond the SM.

The $B$ decays that can yield {\it clean} information about the angles in
the unitarity triangles are, for the most part, decays of the neutral $B$
mesons $B_d (\bar bd)$ and $B_s (\bar bs)$. The physics of the $B_s
-\overline{B_s}$ system is similar to that of the $B_d -\overline{B_d}$
system,\cite{3} so we shall discuss only the latter. The key feature of the
$B_d -\overline{B_d}$ system is that the $B_d (\bar bd)$ and the
$\overline{B_d} (b\bar d)$ mix. In the SM, this mixing is due largely to
the WW box diagram in Fig.~2. The $B_d\rightarrow \overline{B_d}$ amplitude
induced by this higher-order diagram is a suppressed one, so that $B_d
\rightarrow \overline{B_d}$ mixing mechanisms from beyond the SM could
conceivably compete with or even dominate over the SM diagram. Thus, the
modication of $B-\bar B$ mixing is perhaps the most promising route through
which non-SM physics could modify CP violation in the $B$ system.

The physics of the $B_d -\overline{B_d}$ system is well-known. However, it
is an important  background to the quantum mechanical discussion of the
next Section, so we shall briefly review it. The $B_d -\overline{B_d}$
system has two mass eigenstates, $B$-heavy $(B_H)$ and $B$-light $(B_L)$,
given by
\begin{equation}
\mid B_{H(L)} \rangle\;\; =\frac{1}{\sqrt{2}} \left[\mid B_d\;\rangle \;
\stackrel{\textstyle +}{(-)}\;
\omega_{Mix} \mid \overline{B_d} \rangle\right] \;.
\end{equation}
Here,  $\omega_{Mix} \equiv \bigg [A\left(B_d \rightarrow \overline{B_d}
\right)/A \left(\overline{B_d} \rightarrow B_d \right)\bigg
]^{\frac{1}{2}},$ where, here and hereafter, we use the letter $A$ to
denote an amplitude. Empirically, $|\omega_{Mix}|$ is known to be very
close to unity, so that $\omega_{Mix}$ is just a phase factor. Thus, only
the {\it phase} of $B-\bar B$ mixing affects $B_H$ and $B_L$, and, through
them, CP violation in neutral $B$ decay.

In the $SM,A(B_d\rightarrow \overline{B_d})$ is given by the box diagram in
Fig.~2. The amplitude $A(\overline{B_d}\rightarrow B_d)$ is then given by
the same box  diagram, but with every quark (antiquark) replaced by an
antiquark (quark). This  replacement has the effect that every CKM factor
in $A(B_d\rightarrow\overline{B_d} )$ is replaced by its complex conjugate
in  $A(\overline{B_d}\rightarrow B_d)$ Thus,
\begin{equation}
\omega_{Mix} = \frac{V_{td}V^*_{tb}}{V^*_{td}V_{tb}}\equiv e
^{-2i\delta^{Mix}
_{CKM}}\;\;,
\end{equation}
where $\delta^{Mix}_{CKM}$ is the ``$B_d-\overline{B_d}$ mixing
phase."

We shall write the masses of $B_H$ and $B_L$ as
\begin{equation}
m_{H(L)}= m \stackrel{\textstyle +}{(-)}\frac{\Delta
m}{2}-i\frac{\Gamma}{2}\;\;,
\end{equation}
where $m$ is their average mass, $\Delta m$ is their mass difference, and
$\Gamma$ is their  width, which they are expected to have in common.

Suppose that at a time $t=0$ a free neutral $B$ is a pure $|B_d >.$ Due to
the mixing, at a later time  $t$ it will no longer be a pure $|B_d >$, but
will have evolved into a state $|B_d(t) >$ which is a  superposition of
$|B_d >$ and $\mid\overline{B_d}>$ given by
\begin{equation}
\mid B_d(t) >\;\;= e^{-i(m-i\frac{\Gamma}{2})t}\{c\mid B_d>-
ie^{-2i\delta^{Mix}_{CKM}}\;s
\mid
\overline{B_d}>\}\;\;.
\end{equation}
Here, $c\equiv\cos (\frac{\Delta m}{2} t)$ and $s\equiv \sin (\frac{\Delta
m}{2} t).$ Note from Eq.~(7) that, until it decays into some final state,
the $B$ which at  $t=0$ was a pure $|B_d>$ oscillates back and forth
between being a pure  $|B_d>$ and a pure $\mid\overline{B_d}>.$ This
behavior will be important in the discussion  of quantum mechanics in the
next Section.

The decay $B_d(t)\rightarrow f$ of the time-evolved particle $B_d(t)$ into
some final state $f$ has a time-dependent rate, $\Gamma_f(t)$, which from
Eq.~(7) is given by
\begin{equation}
\Gamma _f(t) = \mid <f\mid T\mid B_d(t)>\mid^2= e^{-\Gamma t}
 \mid c<f\mid T\mid B_d>-\;ie^{-2i\delta^{Mix}_{CKM}} s<f\mid
T\mid\overline{B_d} >\mid^2\;\;.
\end{equation}
Let us assume that in this expression, the amplitude $<f|T|B_d >$ for  the
pure  $|B_d>$ decay is dominated by a single Feynman diagram. This
assumption, which is  expected to be a good one for some of the more
important decay modes, is essentially  the only assumption that the
standard analysis of neutral $B$ decays entails. With  this assumption, we
may write
\begin{equation}
<f\mid T\mid B_d>\;\;= Me^{i\delta ^{f}_{CKM}} e^{i\alpha _{ST}}
\end{equation}
Here, $M$ is the magnitude of the dominating diagram, $\delta ^f_{CKM}$ is
the phase of  the product of CKM elements to which this diagram is
proportional, and $\alpha_{ST}$  is a phase due to strong-interaction
effects. Suppose, now, that $f$ is a CP eigenstate, so that  $CP|f > =
\eta_f |f >$, with $\eta_f$ the CP parity of $f$. Then
\begin{eqnarray}
< f\mid T \mid \overline{B_d}> & = & \eta_f<CP[f]\mid T\mid
CP[B_d]>\nonumber \\
 & = & \eta_fMe^{-i\delta^f_{CKM}}e ^{i\alpha _{ST}}\;\;.
\end{eqnarray}
Here, we have used the fact that amplitudes for CP-mirror-image processes,
such  as\linebreak
$<f|T|B_d>$ and $<CP [f]|T| CP[B_d]>$, have opposite CKM phase. This  is
due to the circumstance that every quark in a process is replaced by its
antiquark in the CP- mirror-image process, so that, as previously
mentioned, every CKM factor is replaced by its  complex conjugate. We have
also used the fact that, apart from CKM phases, SM amplitudes are CP
invariant, so that $<f|T|B_d>$ and $<CP[f]|T|CP[B_d]>$ have the same
magnitude and strong phase. From Eqs.~(8),(9), and (10), we have
\begin{equation}
\Gamma _f(t)=M^2e^{-\Gamma t}\bigg\{1-\eta_f\sin\phi\sin(\Delta m
t)\bigg\}\;\;,
\end{equation}
with
\begin{equation}
\phi \equiv 2(\delta^{Mix}_{CKM} + \delta^{f}_{CKM})\;\;.
\end{equation}

When $f$ is a CP eigenstate, the CP mirror image of $B_d(t)\rightarrow f$
is $\overline{B_d}(t) \rightarrow f,$ where the time--evolved state
$\overline{B_d}(t)$ was a pure $\overline{B_d}$ at $t=0$.  Since SM
amplitudes for CP-mirror-image processes are identical  except for having
opposite CKM phases, Eq.~(11) implies that the rate for
$\overline{B_d}(t)\rightarrow f, \bar \Gamma_f(t)$, is given by
\begin{equation}
\bar\Gamma_f(t) = M^2e^{-\Gamma t}\bigg\{1 + \eta_f\sin\phi\sin(\Delta
mt)\bigg\}\;\;.
\end{equation}
The CP-violating asymmetry between $\bar \Gamma_f(t)$ and $\Gamma_f(t)$ is
then
\begin{equation}
\frac{\bar\Gamma_f(t)-\Gamma_f(t)}{\bar\Gamma_f(t) + \Gamma_f(t)}=
\eta_f\sin\phi\sin(\Delta mt)\;\;.
\end{equation}
The mass difference $\Delta m$ is known (at least for the $B_d$ system),
and  $\eta_f$ will be known for  any chosen final state $f$, so a
measurement of this asymmetry will cleanly  determine $\sin\phi$.\cite{4}
Note from Eq.~(12) that the angle $\phi$ that is determined in this way is,
as previously  stated, the phase of a product of CKM elements.

While the case where $f$ is a CP eigenstate is, both theoretically and
experimentally,  the simplest one, clean information on the phases of
products of CKM elements can also be extracted  from many hadronic
$\stackrel{(-)}{B_d}(t)$ and  $\stackrel{(-)}{B_s}(t)$ decays where $f$ is
not a CP eigenstate.\cite{5}  The angle $\phi$ determined by decay into a
final state which is not a CP  eigenstate is no longer given by Eq.~(12),
but is still the relative CKM phase of the two  interfering amplitudes in
the expression (8) for the decay rate.

In Table 1, we list some decay modes that are being considered as possible
probes of  the various angles in the unitarity triangles.  The last column
of this Table gives the angle that can be cleanly determined via study of
each  mode. The one charged B decay listed in the Table illustrates the
fact that  occasionally even charged B decays can provide clean CKM phase
information.

Experiments based on the decay modes of Table 1 and others will be carried
out both at dedicated  high--luminosity $e^+e^-$ colliders(``B
factories")$\cite{11}$ and at hadron  facilities.$\cite{12}$ These
experiments, with their differing strong points, should prove to be quite
complementary.

How could physics beyond the SM (PBSM) affect CP violation in $B$ decays?
As we have seen, $B-\bar B$ mixing, being  suppressed in the SM, is perhaps
the ingredient of CP violation in the $B$ system most susceptible to  the
effects of PBSM. Let us briefly mention three examples of non-SM physics
which,  conceivably, could modify CP violation in $B$ decays by altering
$B-\bar B$ mixing.
\begin{enumerate}
\item Suppose that, in addition to the three known quark $SU(2)_L$ doublets,
there is  also a charge -1/3 singlet.\cite{13} Then, the $Z$ boson can have
quark couplings  which, in mass-eigenstate basis, are nondiagonal. These
nondiagonal couplings can carry  phases beyond those in the 3 x 3 CKM
matrix of the SM. Thus, one can have at tree level the process $\bar
bd\rightarrow Z \rightarrow \bar db,$ and this non-SM  contribution to
$B_d- \overline{B_d}$ mixing can carry a non - SM phase. By modifying the
phase of $B_d-\overline{B_d}$ mixing, this contribution can modify CP
violation in  $\stackrel{(-)}{B_d}(t)$ decay.
\item Suppose that nature contains not just one Higgs multiplet, as in the
$SM,$ but several. Then there can be spontaneous CP violation - a condition
in which different neutral Higgs fields develop vacuum expectation  values
which,  relative to one another, are not real. New contributions to $B-\bar
B$ mixing can include  tree-level processes of the type
$\bar{b}d\rightarrow  H\rightarrow\bar{d}b,$ where  $H$ is some neutral
Higgs boson. The spontaneous CP violation can  impart to these processes
nontrivial, non-SM phases. Then, the phase of $B-\bar B$ mixing, and
consequently CP violation in  neutral $B$ decay, is altered.\cite{14}
\item Suppose that the world is described by some version of supersymmetry
(SUSY). In non-minimal SUSY models,  there can be new phases beyond those
in the CKM matrix, and CP violation in the $B$ system can be altered
substantially.\cite{15} However, there are also minimal SUSY models in
which the avoidance of potentially large flavor  changing neutral currents
has the consequence that there are {\it no} extra phases beyond those in
the CKM matrix. Now, these models do contain non-SM contributions to
$B-\bar B$ mixing, such as the  gluino-squark box diagram in Fig.~3.
Naively, one might imagine that these extra contributions have a different
dependence on the CKM phases than does the SM box diagram of Fig.~2, so
that the  phase of $B-\bar B$ mixing will differ from its SM value.
However, in reality the CKM phases of  the extra contributions are to a
very good approximation identical to the CKM phase of the SM box diagram,
so that the phase of $B-\bar B$ mixing is {\it the same} as in the
SM.\cite{16,15} Now, we have seen that it is only the phase of the $B-\bar
B$ mixing amplitudes, and not their  magnitudes, that influences CP
violation. Thus, CP violation in neutral $B$ decay will also be the same as
in the SM.
\end{enumerate}

It is interesting to ask whether the SUSY contributions to $B-\bar  B$
mixing, while not affecting  CP violation, could still be uncovered by
overconstraining the $B$ system. It is estimated that these contributions
could change the magnitudes of the mixing amplitudes, and consequently the
neutral $B$ mass  differences $\Delta m,$ by (10-20)\%.\cite{17,18} In the
SM, the difference $\Delta m_d$ between the masses of the mass eigenstates
of the $B_d-\overline{B_d}$ system arises from the box diagram in Fig.~2,
and so is proportional to  $|V_{td}|^2$. The analogous mass difference
$\Delta m_s$ in the $B_s-\overline{B_s}$ system arises from a  similar
diagram in which the $d$ quarks have been replaced by $s$ quarks, and so is
proportional to  $|V_{ts}|^2.$ Thus, in the SM,
\begin{equation}
\sqrt{\frac{\Delta m_d}{\Delta m_s}} = C\frac{\mid V_{td}\mid}
{\mid V_{ts}\mid}\;\;,
\end{equation}
where the coefficient $C$ is expected to be  approximately unity, and, more
precisely, is estimated  to be 0.86 $\pm$ 0.1.\cite{19} Now, we know from
unitarity that  $|V_{ts}| \cong |V_{cb}|$, and $|V_{cb}|$ is known to
$\sim$ 15\%. Thus, if $B-\bar{B}$ mixing comes only from the  SM box
diagram, a measurement of $\Delta m_d/\Delta m_s$ would  determine
$|V_{td}|$. Since $| V_{tb}| \cong 1$, we would then know the length of the
$V_{td}V_{tb}$ side of the unitarity triangle of  Fig.~1. If we had also
learned, from other sources, the lengths of the other two sides, we could
then deduce the  interior angles.

Now, if $B-\bar B$ mixing actually contains significant non-SM
contributions from  SUSY, then one might expect the SM relation (15) to
fail. The length  $|V_{td}V_{tb}|\cong |V_{td}|$ of the $V_{td}V_{tb}$ side
of the unitarity triangle deduced from $\Delta m_d/\Delta m_s$ by using the
relation (15) would then  be wrong. Thus, the interior angles of the
triangle inferred from the alleged lengths of the sides  would be wrong as
well, and would disagree with the true interior angles  determined by
measurements  of CP - violating asymmetries. This disagreement would be the
signal of physics beyond the SM.

Unfortunately, it appears that the SUSY correction to $\Delta m_q(q=d$  or
$s$) is proportional to its SM value. Thus, although SUSY may change
$\Delta m_d$ and $\Delta m_s$ individually by $\sim$ 20\%,  it does not
visibly alter their ratio from its SM size.\cite{17,20} Hence,unlike the
other models we mentioned, this minimal version of SUSY is  an interesting
example of non-SM physics whose presence could not be detected by studies
of CP violation or mixing in neutral $B$ decays.

\section*{Quantum Mechanics at $B$ Factories}

The studies of CP violation to be performed at $B$ factories will involve
some very interesting quantum  mechanics, to which we now turn.\cite{21}

At the B factories, the $B$ mesons whose decays are to be studied will be
produced via the reaction  $e^+e^-\rightarrow\Upsilon(4S)\rightarrow BB$.
The $\Upsilon(4S)$ is a $\bar bb$ bound state which  decays into $B$ pairs
essentially 100\% of the time. Roughly half of its decays yield $B^+B^-,$
and  the other half $B_d\overline{B_d}$. (The $\Upsilon(4S)$ is not heavy
enough to decay to $B_s\bar B_s$.)

Since it is mostly the decays of neutral $B$ mesons which can yield clean
CKM phase information,  we shall be interested in events where
$\Upsilon(4S)\rightarrow B_d\overline{B_d}.$ Now, the  $\Upsilon(4S)$ has
spin of unity, and $B$ mesons are spinless, so the $B$ pair produced by
$\Upsilon(4S)\rightarrow B_d\overline{B_d}$ is in a {\it p} wave . Let us
view this pair in the $\Upsilon(4S)$ rest frame, where the $B$ mesons are
moving outward in opposite directions from the $\Upsilon(4S)$ decay point.
Due to $B_d-\overline{B_d}$ mixing,  each of the two $B$ mesons is
oscillating back and forth between pure $B_d$ and pure $\overline{B_d}$
(see Eq.~(7)).  However, at no time may we have two identical spinless
bosons in a {\it  p} wave. Thus, if at some time $t$ one of the $B$ mesons
is found to be, say, a $\overline{B_d},$ then at this time the other  $B$
meson must be a $B_d$. This is a modern example of an
Einstein-Podolsky-Rosen (EPR) correlation.\cite{22}

This EPR correlation plays a crucial role in the traditional
description\cite{23} of a typical $B$ factory experiment.  Let us recall
this description. The sequence of events in the experiment, viewed from the
$\Upsilon(4S)$ rest frame, is shown in Fig.~4. At a time we shall call
$t=0$, the  $\Upsilon(4S)$  decays into a pair of neutral B mesons, which
move outward back to back. At a subsequent  time $t_\ell$, one of the $B$
mesons decays semileptonically, and we shall suppose that it yields, in
particular, a negatively-charged lepton $\ell^-$, plus other particles
$X.$   At another time $t_{CP}$, the other $B$ meson decays into a hadronic
CP eigenstate, which we shall take for  illustration to be $\pi^+\pi^-$.

Let us suppose first that the semileptonic decay occurs before the one to
$\pi\pi:t_\ell <t_{CP}.$    Now, only a $\overline{B_d},$ but not a $B_d,$
can decay to  a {\it negative} lepton, so the charge of the $\ell^-$ in
Fig.~4 indentifies its parent as being a $\overline{B_d}$ at the instant of
decay, $t_\ell.$ Since, at any given time, one cannot have two identical
bosons in a  {\it p} wave,  this means that {\it at the same instant
$t_\ell$} the other $B$, on the right in Fig.~4, must be a pure $B_d$. The
decay of one $B$ into an $\ell^-$ has  ``collapsed the wave function" for
the $BB$ state, leaving behind a  single $B$ whose state, at the time of
collapse, is known precisely.  Of course, subsequent to the time $t_\ell$,
the surviving $B$ will oscillate  between pure $B_d$ and pure
$\overline{B_d}$ because of mixing.  Taking advantage of the fact that, in
the $\Upsilon(4S)$ rest frame,  the $B$ mesons are rather nonrelativistic
$(\frac{v}{c}\sim 0.06)$,  we may neglect their motion. The probability for
the $B$ which  survives beyond $t_\ell$ to decay into the CP eigenstate
$\pi^+\pi^-$ is then given by the $B$-rest-frame expression (11), in  which
to a very good approximation we need not distinguish between  time in the
$B$ rest frame and in the $\Upsilon(4S)$ frame. However,  we must note that
the time variable $t$ in expression (11) represents  the time of the decay
to the CP eigenstate (here, $t_{CP})$ relative  to the time when the parent
$B$ was known to be pure $B_d$. In the  present case, the latter time is
not $t=0,$ the instant when the $B$  was born, but $t=t_\ell$, the instant
when the {\it other} $B$ decayed  to an $\ell^-$. That is, in applying
Eq.~(11), we must take  $t=t_{CP}-t_\ell$. The joint probability for one
$B$ to decay to  $\ell^-X$ at time $t_\ell$ and the other to decay to
$\pi^+\pi^-$ at  time $t_{CP}, \Gamma \bigg[{\rm One}\;\; B\rightarrow
\ell^- X\;\;{\rm at}\;\; t_\ell\;  \;;{\rm Other}\;\;  B\rightarrow
\pi^+\pi^- \;\;{\rm at}\;\; t_{CP}\bigg], $ is then given by
\begin{eqnarray}
&& \Gamma  \bigg[{\rm One}\;\; B\rightarrow \ell^- X\;\; {\rm at}\;\;
t_\ell;\;{\rm Other}\;\;
B\rightarrow \pi^+\pi^- \;\;{\rm at}\;\; t_{CP}\bigg]\nonumber \\
&& \propto e^{-\Gamma t_\ell}e^{-\Gamma
t_\ell}e^{-\Gamma(t_{CP}-t_\ell)}\{ 1-\sin\phi\sin\left[\Delta
m(t_{CP}-t_\ell)\right]\} \\
&& = e^{-\Gamma (t_{CP} + t_\ell)}\bigg\{1-\sin\phi\sin\left[\Delta
m(t_{CP}-t_\ell)\right]\bigg\}\;\;. \nonumber
\end{eqnarray}
In the first part of Eq.~(16), the first factor, $e^{-\Gamma t_\ell},$ is
the probability for the $B$ which yields $\ell^-X$ to survive  until time
$t_\ell$; the second factor, $e^{-\Gamma t_\ell},$ is the  probability for
the $B$ which will eventually yield $\pi^+\pi^-$ to  survive until time
$t_\ell$; and the remaining factors are, from  Eq.~(11), the probability
for the latter $B$ to decay to $\pi^+\pi^-$,  given that at time $t_\ell$
it was a pure $B_d$. In writing Eq.~(16),  we have used the fact that for
$\pi^+\pi^-, \eta_f=+1.$

What if the decay to $\pi^+\pi^-$ occurs before the semileptonic one
$(t_{CP}<t_\ell$)? To answer this question, we note that there is  always a
linear combination of $|B_d>$ and $\mid\overline{B_d}>,  |B_{no}>\linebreak
\propto <\pi^+\pi^-|T\mid \overline{B_d}>|  B_d>-<\pi^+\pi^-|T|B_d>\mid
\overline{B_d}>$, which has  vanishing amplitude for decay to $\pi^+\pi^-.$
Now, it is obvious that  if the $B$ meson on the right in Fig.~4 decays to
$\pi^+\pi^-$, then  at the time of its decay, $t_{CP},$ it is not in the
state $|B_{no}>.$  But then, since one cannot have two identical bosons in
a {\it p} wave, at  this same time, $t_{CP},$ the $B$ meson on the left in
Fig.~4 {\it is  }in the state $|B_{no}>$. More generally, if $\Upsilon(4S)
\rightarrow BB$ (where the $B$ mesons are neutral), and one of the $B$
mesons decays to some final state $f$, then the other $B$ meson cannot
decay to the same final state {\it at the same time}.\cite{24,25}  Thus,
the decay of one $B$ to $\pi^+\pi^-$ ``collapses the BB wave  function" and
fixes the state of the remaining $B$ as $|B_{no}>$  at time $t_{CP}$. The
state of this remaining $B$ may then be evolved  forward in time from the
instant $t_{CP}$ using the Schr\"{o}dinger  equation for the
$B_d-\overline{B_d}$ system, and one may calculate  the amplitude for the
time-evolved state to decay to $\ell^-X$ at time  $t_\ell$. Interestingly,
one finds that the joint probability for one  $B$ to decay to $\pi^-\pi^-$
at time $t_{CP}$, and the other to   $\ell^-X$ at time $t_\ell > t_{CP},$
is given by the same  expression as before, Eq.~(16). This equation holds
regardless of the  order of events.\cite{26}

Having reviewed the traditional description of this typical B-factory
experiment, let us ask how our picture of  the experiment is  modified if
we require consistency with relativity and take the motion  of the $B$
mesons fully into account. When relativity is not neglected,  several
issues arise: First, in relativity, the simultaneity of two events depends
on the frame  of reference.   Thus, if in the traditional treatment one
asserts that the decay of one $B$ fixes the state of the other $B$ {\it  at
the same time,} one must specify in which frame of reference this
assertion is true. Which frame is it, and why? Secondly, in some events,
the separation between the two $B$ decays in  Fig.~4 will be spacelike.
Then, which $B$ decays first, collapsing the  wave function and fixing the
state  of the other $B$, will depend on  the  frame of reference. That is,
in the traditional treatment,    the same collection of occurences will be
described quite differently in  different frames. It would appear that an
alternative treatment which  is not so strongly frame dependent would be
advantageous. Thirdly, what are the relativistic  corrections to Eq.~(16),
the  expression which will be used to extract the values of CKM phases
$\phi$ from B-factory experiments? \def\subdecay#1{\raise 5pt \hbox{\hspace
{0.1 em} {\vrule height 15pt depth -2.8 pt } \hspace {-0.75 em}
$\longrightarrow #1$} }

Let us generalize the decay sequence of Fig.~4 to include any chain of
the form,
\begin{eqnarray}
\Upsilon(4S)\rightarrow  B & + & B\;\;\;\;\;\;\;\;\;\;, \\
&\!\!\! \!\!\!\!\!\subdecay{f_1(t_1,\vec x _1)} &
\subdecay{f_2(t_2,\vec x_2)}\nonumber
\end{eqnarray}
where the $B$ mesons are neutral, the $f_j$ are arbitrary final  states,
and $(t_j,\vec{x_j})$ is the spacetime point where the decay to $f_j$
occurs.  To take relativity into account and address the issues just
raised,  let us treat any chain of this type by directly calculating the
amplitude for it, \cite{21} without introducing the wave function for  the
$BB$ state, or invoking the collapse of this wave function. To  calculate
the amplitude for (17), it is convenient to work in the $B$  mass
eigenstate basis. The amplitude has two terms. One of these  describes the
process in which the $B$ meson which decays to $f_1$ is  a $B_H$, while the
one which decays to $f_2$ is a $B_L$. The other  term describes a process
in which the roles of $B_H$ and $B_L$ are  interchanged. Since the
$B_H-B_L$ mass difference is tiny, these two  processes are experimentally
indistinguishable, so their amplitudes  must be added coherently. Owing to
the antisymmetry of the amplitude  for $\Upsilon(4S)\rightarrow BB,$ it is
not possible for both $B$  mesons to be $B_H$ or $B_L$.

The amplitude $A_{HL}$ for the process where it is $B_H$ which decays
to $f_1$ and $B_L$ which decays to $f_2$ is given by
\begin{equation}
A_{HL} = A(B_H\;\; {\rm to} \;\;1;B_L \;\;{\rm to} \;\;2)e ^{
-im_H\tau_1}e^{-im_L\tau_2}A(B_H\rightarrow f_1)A(B_L\rightarrow f_2)\;\;.
\end{equation}
Here, $A(B_H\;\; {\rm to}\;\; 1; B_L \;\;{\rm to} \;\;2)$ is the amplitude
for an $\Upsilon(4S)$  to decay to a $B_H$ moving towards the point
$(t_1,\vec{x_1})$ and a  $B_L$ moving towards $(t_2,\vec{x_2}).$  The  only
feature of this amplitude which will be relevant is its  antisymmetry under
$B_H\leftrightarrow B_L.$ The factor $\exp [-im_H\tau_1]$ is the amplitude
for the $B_H,$ which  has complex mass $m_H$(including its width), to
propagate from the  spacetime point where it is produced by the
$\Upsilon(4S)$ decay to  the point ($t_1, \vec{x_1})$ where it decays. In
this factor,  $\tau_1$ is the proper time, in the $B_H$ rest frame, which
elapses  during this propagation. Similarly, the factor $\exp
[-im_L\tau_2]$  is the amplitude for the $B_L$ to propagate to $(t_2,
\vec{x_2})$.  That the amplitude for a particle of mass $M$ to propagate
for a  proper time $\tau$ is $\exp[-iM\tau]$ may be understood by solving
Schr\"{o}dinger's equation for the time evolution of such a particle  in
its rest frame, and then noticing that the solution, $\exp  [-iM\tau]$, is
Lorentz-invariant.\cite{27} Finally, the factor  $A(B_H\rightarrow f_1)$ in
$A_{HL}$ is the amplitude for $B_H$ to  decay to $f_1$, and similarly for
$A(B_L\rightarrow f_2).$ If the various A's on the right-hand side of
Eq.~(18) are invariants,  $A_{HL}$ is an invariant.

Adding to the $A_{HL}$ of Eq.~(18) the analogous expression for
$A_{LH}\equiv A_{HL} (B_H\leftrightarrow B_L$), and taking into  account
the antisymmetry of  $A(B_H\;\;{\rm to}\;\; 1; B_L\;\;{\rm to}\;\;2)$ under
$B_H\leftrightarrow B_L,$ we find that  the complete amplitude $A$ for the
decay chain (17) is given by
\begin{eqnarray}
A & \propto & e^{
-im_H\tau_1}e^{-im_L\tau_2}A(B_H\rightarrow f_1)A(B_L\rightarrow
f_2)\nonumber \\
&&   -e^{
-im_L\tau_1}e^{-im_H\tau_2}A(B_L\rightarrow f_1)A(B_H\rightarrow
f_2)\;\;.
\end{eqnarray}

Obviously, the amplitude approach which has yielded this simple result
 is quite general. It can be applied, for example, to multibody
sequences of the form
\begin{equation}
\begin{array}{llll}
P\rightarrow \!\!  & A\;\;\;\;   + &     B  \;\;\;\;  + &    C  \;\;\;\;  +
\cdots\;\;,\nonumber\\
& \!\subdecay{f_1} & \!\subdecay{f_2}  & \!\subdecay{f_3} \nonumber
\end{array}
\end{equation}
in which $P$ can be a single particle or two particles which have
collided, and the ``decays" $A\rightarrow f_1,$ etc., can alternatively
be measurements of various properties of the particles A,B,...

In the traditional collapsing wave function description of decay  sequences
of the type (17), one invokes the fact that if one of the  $B$ mesons
decays to some final state $f$ at a time $t,$ then the other  $B$ cannot
decay to the same final state {\it at the same time}. This  fixes the state
of the surviving $B$ at time $t.$ However, if  relativity is not neglected,
then, as we have noted, one must ask {\it  in which Lorentz frame} the
simulaneous decay of the two $B$ mesons to  the same final state is
impossible, so that the decay of the one fixes  the state of the other. The
amplitude (19) answers this question. For,  if we take $f_1 =f_2$, then
this amplitude vanishes for  $\tau_1=\tau_2$. That is, the two $B$ mesons
cannot decay in the same  way at equal {\it proper} times in their
respective rest frames. This  implies that it is the $\Upsilon(4S)$ rest
frame in which they cannot  decay to the same final state simultaneously.
For, in the $\Upsilon  (4S)$ frame the two $B$ mesons have equal speed, so
that, after time  dilation, equal proper times in the two $B$ rest frames
correspond to  a single, common time in the $\Upsilon (4S)$ frame.
\newpage
Let us now apply our general amplitude (19) to the specific decay
chain of most interest to $B$ factories,
\begin{eqnarray}
\Upsilon(4S)\rightarrow  B & + & B\;\;\;\;\;\;\;\;\;\;\;\;,\\
& \!\!\!\!\!\!\!\!\!\!\subdecay{\ell^-X}  & \subdecay{f_{CP}}\nonumber
\end{eqnarray}
where $f_{CP}$ is a   CP eigenstate. Taking $f_1=\ell^-X$ in
(19) and using Eqs.(4) and (5) and the fact that only a
$\overline{B_d}$, but not a $B_d$, can decay to $\ell^-X$, we have
\begin{eqnarray}
A(B_{H(L)}\rightarrow f_1) & = & < \ell^-X\mid T\mid
B_{H(L)}>\nonumber\\
=\stackrel{\textstyle +}{(-)}\frac{1}{\sqrt{2}} \!\! & \!\! e \!\! & \!\!
\!\!^{-2i\delta^{Mix}_{CKM}}<\ell^-X\mid
T\mid \overline{B_d}>\;\;.
\end{eqnarray}
Taking $f_2=f_{CP}$, assuming as before that one diagram dominates $B
_d\rightarrow f_{CP},$ and using Eqs.~(4), (5), (9), and (10), we have
$$
A(B_{H(L)}\rightarrow f_2)=<f_{CP}\mid T\mid B_{H(L)}>
$$
\begin{equation}
=\frac{1}{\sqrt{2}}Me^{i\delta^{f}_{CKM}}e^{i\alpha_{ST}}
(1\stackrel{\textstyle +}{(-)}\eta_f e^{-i\phi})\;\;.
\end{equation}
Here, $\eta_f$ is the CP parity of the final state, and the CKM
phase $\phi \equiv 2(\delta^{Mix}_{CKM} + \delta ^f_{CKM}),$ as
before, (cf. Eq.~(12)). Let us write the proper time $\tau_1$ of the
decay to $\ell^-X$ as $\tau _\ell$, the proper time $\tau_2$ of the
decay to $f_{CP}$ as $\tau_{CP},$ and the masses $m_{H(L)}$ of $B_{H(L)}$
as in Eq.~(6). Then, omitting irrelevant overall factors, we find from
Eqs.~(19), (22), and (23) that the amplitude $A$ for the decay chain (21)
is given by
\begin{eqnarray}
A &\propto & e^{-\frac{\Gamma}{2}(\tau_{CP}+\tau_\ell)}\{\cos[\frac{\Delta
m}{2}(\tau_{CP}-\tau_\ell)]\nonumber \\
&& -i\eta_f e^{-i\phi}\sin[\frac{\Delta
m}{2}(\tau_{CP}-\tau_\ell)]\}\;\;.
\end{eqnarray}
Taking the absolute square of this expression, we find for the joint
probability for one $B$ to decay to $\ell^-X$ at proper time
$\tau_\ell$,
and the other to decay to $f_{CP}$ at proper time $\tau_{CP},$
\begin{eqnarray}
& \Gamma & \!\![{\rm One} \;\;B\rightarrow \ell^-X \;\;{\rm at}\;\;
\tau_\ell;
\;\;{\rm Other}\;\; B\rightarrow
f_{CP}\;\; {\rm at}\;\; \tau_{CP}]\nonumber \\
& \propto &  e^{-\Gamma (\tau_{CP}+\tau_\ell)}\{1-\eta_f\sin\phi\sin[\Delta
m(\tau_{CP}-\tau_\ell)]\}\;\;.
\end{eqnarray}
This result agrees with the one of Eq.~(16), obtained by collapsing  the
$BB$ wave function. (The factor $\eta_f$ is absent from Eq.~(16)  because
that expression was derived for the illustrative case where  $f_{CP} =
\pi^+\pi^-,$ a {CP} eigenstate with $\eta_f= +1.)$ However, the
relativistically precise Eq.~(25) makes it clear that if the  expression
(16) for the joint decay probability is to be totally  accurate, then the
$B$ decay times in it must be taken to be {\it  proper} times, not times in
the $\Upsilon (4S)$ rest frame. This is a negligible distinction for
$\Upsilon (4S)\rightarrow BB$, but for the  similar process $\phi
\rightarrow KK\rightarrow  (\pi^+\pi^-)(\pi^\circ\pi^\circ)$,\cite{25} to
be  studied at the $\phi$  factory DA$\Phi$NE,\cite{28} use of $\phi$-frame
times rather than $K$-frame   proper times would cause a (2-3)\% error. It
is possible to refine the  ``wave function collapse" approach so that it
takes the motion of the  $B$ mesons properly into account and yields
exactly the same result,  Eq.~(25), as the ``amplitude" approach, with no
ambiguity about the  time variables.\cite{29} However, we\cite{29} would
not have known how to do   so without using the ``amplitude" result for
guidance.

The amplitude approach has also been applied to $\Upsilon (4S)
\rightarrow BB\rightarrow (\ell X) (f),$ where $f$ is not a CP
eigenstate, and to processes where the $B$ meson pair is in a
symmetric state, rather than the antisymmetric one of $\Upsilon (4S)$
decay.\cite{29} These applications, like the one to $\Upsilon(4S)
\rightarrow BB\rightarrow (\ell X)(f_{CP})$ which we have discussed,
confirm the results of the collapse approach, while taking relativity
fully into account, and avoiding the puzzles associated with collapse
of the wave function.
\section*{Summary}

Study of ${CP}$ violation in $B$ decay will provide a powerful
test of the SM of CP violation, and a probe of physics beyond
the SM. Both to test, and to look for physics beyond, the SM, it
will be very important to overconstrain the $B$ system as much as
possible, through measurement of a variety of quantities at different
experimental facilities. The measurements to be carried out at the
$e^+e^-\; B$ factories will involve some very pretty quantum mechanics,
which can be simply understood via an amplitude approach which does
not entail the ``collapse of the wave function."

\section*{Acknowledgements}
It is a pleasure to thank A.~Ali, F.~Borzumati, G.~Giudice,
A.~Masiero, and R.~Mohapatra for informative conversations. I would
also like to thank the Deutsches Elektronen-Synchrotron (DESY) and the
Fermi National Accelerator Laboratory for their warm hospitality
during the time that this talk and its written version were prepared.

%


\begin{table}[hb]
\caption{Decay modes and the phase angle $\phi$ which they
probe. In the final state $\psi K^{*\circ},$ the $K^{*\circ}$ is
required to decay as shown. Similarly for the final state
$\stackrel{(-)}{D^\circ} K^+; g_{CP}$ is a CP eigenstate, such a
$\pi^+\pi^-$ or $K^+K^-.$ References are given in the first column.}
\begin{center}
\begin{tabular}{|l|l|l|}\hline
{Ref.} & $\;\;\;\;\; $ Decay Mode & $\phi$ \\
\hline
\hline
4,6 & $B_d(t)\rightarrow \pi^+\pi^-, \pi^+\rho^-,\pi^+{\rm a}_1^-$ &
2$\alpha$ \\
4,7 &  $B_d(t)\rightarrow \psi K_s, \psi K^{*\circ}$ & 2$\beta$ \\
{} & $\;\;\;\;\;\;\;\;\;\;\;\;\;\;\;\;\;\;\;\;\;\;\;\;\;
\subdecay{K_s\pi^\circ}$ & {} \\
8 & $B_s (t)\rightarrow D^+_s K^-$ & $\gamma$ \\
9 & $B^+\rightarrow \stackrel{(-)}{D^\circ}K^+ $ & $\gamma$ \\
{} & $\;\;\;\;\;\;\;\;\;\;\subdecay{g_{CP}}$ & {} \\
10 & $B_s(t)\rightarrow \psi\phi$ & 2$\epsilon$ \\
\hline
\hline
\end{tabular}
\end{center}
\end{table}

\begin{figure}[ht]
\caption{The $db$ unitarity triangle.}
\caption{The SM diagram for $B_d\rightarrow\overline{B_d}$ mixing.}
\caption{A non-SM contribution to $B_d\rightarrow \overline{B_d}$ mixing in
SUSY. The $\tilde g$
is a gluino, $\tilde d$ and $\tilde b$ are squarks, and the symbol X stands
for squark mixing.}
\caption{The events in a typical B factory experiment,
viewed in the $\Upsilon(4S)$ rest frame.}
\end{figure}

\begin{thebibliography}{99}
\bibitem{1} R. Aleksan, B.~Kayser, and D.~London, {\it Phys.
Rev.~Lett.} {\bf 73}, 18 (1994)
\bibitem{2} L. Wolfenstein, {\it Phys.Rev.~Lett.~} {\bf 51}, 1945 (1983).
\bibitem{3} See, however, I.~Dunietz, Fermilab preprint
FERMILAB-PUB-94/361-T.
\bibitem{4} A. Carter and A.~Sanda, {\it Phys.~Rev.~Lett.} {\bf 45}, 952
(1980); {\it Phys. Rev.~D} {\bf 23}, 1567 (1981); I.~Bigi and A. Sanda,
{\it Nucl.~Phys.~B}{\bf 193}, 85 (1981);{\bf 281}, 41 (1987);
L.~Wolfenstein, {\it Nucl.~Phys.~B} {\bf  246}, 45 (1984); and
I.~Dunietz and J.~Rosner, {\it Phys. Rev.~D} {\bf 34}, 1404 (1986).
\bibitem{5} B. Kayser, M.Kuroda, R. Peccei, and A. Sanda, {\it Phys.
Lett.~B} {\bf 237} 508 (1990); I. Dunietz, H. Lipkin, H. Quinn, A.
Snyder, and W. Toki, {\it Phys.~Rev.~D} {\bf 43}, 2193 (1991); R.
Aleksan, I Dunietz, B. Kayser and F.~Le Diberder, {\it Nucl.~Phys.~B}
 {\bf 361}, 141 (1991).
\bibitem{6} M.~Gronau and D. London, {\it Phys. Rev.~Lett.} {\bf 65},
3381 (1990); Aleksan et al. Ref.~5; A. Snyder and H. Quinn, {\it Phys. Rev.
D} {\bf 48}, 2139 (1993);F. DeJongh and P. Sphicas, Collider
Detector Facility note CDF/PHYS/BOTTOM/ PUBLIC/3045.
\bibitem{7} B. Kayser et al. and I. Dunietz, et al., Ref. 5.
\bibitem{8} R. Aleksan, I. Dunietz, and B.~Kayser,{\it Zeit.
Phys.~C} {\bf 54}, 653 (1992).
\bibitem{9} M.~Gronau and D. Wyler, {\it Phys. Lett. B} {\bf 265}, 172
(1991).
\bibitem{10} I. Dunietz, {\it Proceedings of the Workshop on B
Physics at Hadron Accelerators}, edited by P.~McBride and C.S. Mishra
(Fermilab, Batavia, IL, 1993) p.~83.
\bibitem{11} Stanford Linear Accelerator Center reports SLAC-418 and
SLAC-R-95-457; KEK (National Lab.~for High Energy Phys.) report
entitled ``KEKB Design Report," to be published in 1995.
\bibitem{12} See the {\it Proceedings of the Workshop on B Physics
at Hadron Accelerators,} edited by P.~McBride and C.S. Mishra (Fermilab
report Fermilab-CONF-93/267).
\bibitem{13} Y.Nir and D. Silverman, {\it Phys. Rev.D} {\bf 42}, 1477
(1990).
\bibitem{14} CP violation in the presence of several Higgs multiplets has
been analyzed recently by Y. Wu and L.~Wolfenstein, {\it Phys. Rev.~Lett.
} {\bf 73}, 1762 (1994).
\bibitem{15} I. Bigi and F. Gabbiani, {\it Nucl.~Phys.~B} {\bf 352}, 309
(1991).
\bibitem{16} T. Kurimoto, Osaka University preprint OS-GE 44-95. I wish to
thank A.~Masiero for an enlightening conversation on the point.
\bibitem{17} S. Bertolini, F. Borzumati, A. Masiero, and G. Ridolfi,{\it
Nucl.  Phys. B} {\bf 353}, 591 (1991).
\bibitem{18} Y. Okada, KEK preprint 94-192.
\bibitem{19} A.Ali and D.London,{\it Proceedings of the 27th
International Conference on High Energy Physics}, edited by P. Bussey and
I. Knowles
(Institute of Physics Publishing, Bristol, 1995) p. 1133.
\bibitem{20} I thank F. Borzumati for a helpful conversation on this
matter.
\bibitem{21} B. Kayser and L. Stodolsky, Max Planck Institut preprint
MPI-PhT/95-47.
\bibitem{22} A. Einstein, B. Podolsky, and N. Rosen, {\it Phys. Rev.}
 {\bf 47}, 777 (1935).
\bibitem{23} S. Rudaz, talk presented at the 1988 Aspen Winter Physics
Conference.
\bibitem{24} This conclusion can, of course, be more rigorously derived.
\bibitem{25} H. Lipkin, {\it Phys. Rev.} {\bf 176}, 1715 (1968). This
paper
has been an inspiration to L. Stodolsky and the present author.
\bibitem{26} R. Blankenbecler, unpublished Stanford Linear Accelerator
Center report.
\bibitem{27} See also L. Stodolsky, {\it Gen. Rel. and Grav.} {\bf 11},
391 (1980).
\bibitem{28} J.~Lee-Franzini, {\it Proceedings of Les Rencontres de
Physique de la Vallee D'Aoste 1992--Results and Perspectives in
Particles Physics}, edited by M.~Greco (Editions Frontieres,
Gif-sur-Yvette, France, 1992) p. 349.
\bibitem{29} B. Kayser and L. Stodolsky, to be reported elsewhere.
\end{thebibliography}
\end{document}